\documentclass[10pt,journal,compsoc]{IEEEtran}
\usepackage{hyperref}

\usepackage{bm}

\usepackage{fontawesome}

\usepackage{tikz}
\usetikzlibrary{shapes,decorations,arrows,calc,arrows.meta,fit,positioning}
\tikzset{
    -Latex,auto,node distance =1 cm and 1 cm,semithick,
    state/.style ={ellipse, draw, minimum width = 0.7 cm},
    point/.style = {circle, draw, inner sep=0.04cm,fill,node contents={}},
    bidirected/.style={Latex-Latex,dashed},
    el/.style = {inner sep=2pt, align=left, sloped}
}

\usepackage{stfloats}

\ifCLASSOPTIONcompsoc{}
  \usepackage[nocompress]{cite}
\else
  \usepackage{cite}
\fi

\usepackage{graphicx}

\ifCLASSINFOpdf{}
\else
\fi

\begin{document}
\title{Software and Dependencies\\in Research Citation Graphs}

\author{Stephan~Druskat
  \IEEEcompsocitemizethanks{\IEEEcompsocthanksitem Stephan Druskat is with the German Aerospace Center (DLR), Berlin, Germany, the Computer Science Department at Humboldt-Universit\"at zu Berlin, Berlin, Germany, and the Department of English Studies at Friedrich Schiller University, Jena, Germany.\protect\\
    E-mail: stephan.druskat@dlr.de}

  \thanks{\textcopyright~2019 IEEE.  Personal use of this material is permitted.  Permission from IEEE must be obtained for all other uses, in any current or future media, including reprinting/republishing this material for advertising or promotional purposes, creating new collective works, for resale or redistribution to servers or lists, or reuse of any copyrighted component of this work in other works. 
  {\tt DOI:} \url{https://doi.org/10.1109/MCSE.2019.2952840}}}

\markboth{Computing in Science and Engineering}%
{Druskat: Software and Dependencies in Research Citation Graphs}
\IEEEtitleabstractindextext{%
  \begin{abstract}
    Following the widespread digitalization of scholarship, software has become essential for research, but the current sociotechnical system of citation
    does not reflect this sufficiently.
    Citation provides context for research, but the current model for the respective research citation graphs does not integrate software.
    In this paper, I develop a directed graph model to alleviate this, describe challenges for its instantiation, and give an outlook of
    useful applications of research citation graphs, including transitive credit.
  \end{abstract}

  \begin{IEEEkeywords}
    software citation, citation graphs, transitive credit
  \end{IEEEkeywords}}

\maketitle

\IEEEdisplaynontitleabstractindextext

\IEEEraisesectionheading{\section{Introduction}\label{sec:introduction}}

\IEEEPARstart{T}{he digitalization} of research changes research methods across disciplines, and produces new forms of research and knowledge~\cite{borgmanScholarshipDigitalAge2007}.
In the process, research software has clearly become an integral part of digital research methodologies~\cite{gobleBetterSoftwareBetter2014}.
It embeds research knowledge, implements algorithms and models, and is a central component of digital scholarly integration and application~\cite{haferAssessingOpenSource2009}. It thus presents a significant, and increasingly vital, intellectual contribution to academic research.

Research software should therefore also be considered a legitimate research product~\cite{haferAssessingOpenSource2009,smithSoftwareCitationPrinciples2016,piwowarAltmetricsValueAll2013}.
Research products have the most value -- and their outcomes can be understood most fully -- when they are considered in their context~\cite[p.~10]{borgmanScholarshipDigitalAge2007}.
This context is standardly provided through citation.
Therefore, one aspect of research software gaining the status of a research product is, that it must be integrated into the scholarly citation system.

The citation system in current digital scholarship is a sociotechnical system based on technical infrastructure, and involves different stakeholders.
Stakeholders include domain-specific communities of researchers who use software,
research software engineers (RSEs) and other software developers, research institutions,
publishers, repository providers, index providers, and funders (cf.~\cite{katzSoftwareCitationImplementation2019}).
Technical infrastructure upon which the citation system is built most prominently include publication repositories; citation indices and aggregators; publishing services including websites; services provided by libraries; resolvers for digital identifiers; metadata formats; reference management, text processing, and other software.

Citation and the sociotechnical citation system broadly provide the following functions:

\begin{itemize}
  \item \textbf{Context function:} The provision of context for research products by establishing a graph of research products with links between the identified citing and cited products, to enable traceability of outcomes, both over the past to understand how present knowledge was established, and into the future to understand how present knowledge is being used (cf.~\cite{debellisBibliometricsCitationAnalysis2009,borgmanScholarshipDigitalAge2007,garfieldCitationIndexesScience1955,katzTransitiveCreditMeans2014})
  \item \textbf{Social functions:} The establishment of trust and authority~\cite{nicholasTrustAuthorityScholarly2014,greenbergHowCitationDistortions2009,katzTransitiveCreditMeans2014}; the recognition of the value of a research product while providing credit to its authors~\cite{smithSoftwareCitationPrinciples2016,debellisBibliometricsCitationAnalysis2009}; the potential for evaluation of individual researchers~\cite[ch.~6]{debellisBibliometricsCitationAnalysis2009}, individual research products~\cite{neylonArticleLevelMetricsEvolution2009}, journals~\cite{garfieldCitationAnalysisTool1972}, and research groups, institutions, and countries~\cite[ch.~6]{debellisBibliometricsCitationAnalysis2009}
  \item \textbf{Compliance function:} The assertion of compliance with good scholarly practice~\cite{debellisBibliometricsCitationAnalysis2009,morinShiningLightBlack2012a}
  \item \textbf{Discursive function:} The organization and shaping of discourses of scholarly credibility, authority, and relevance through epistemic change via ``dynamically rewriting the past''~\cite[p.~xvi]{debellisBibliometricsCitationAnalysis2009},~\cite{bungeEpistemicChange1983,foucaultArchaeologyKnowledge1982,greenbergHowCitationDistortions2009})
  \item \textbf{Reproducibility function:} The enablement of research reproducibility through correct and complete citation~\cite{berez-kroekerReproducibleResearchLinguistics2018,doerrGivingSoftwareIts2019,cousijnDataCitationRoadmap2018,pengReproducibleResearchComputational2011}
\end{itemize}

Software as a research product can be subject to all of the described functions -- including the discursive function, albeit to a limited degree, see below -- only if it is fully integrated in the citation system.
While this is not currently the case~\cite{howisonSoftwareScientificLiterature2016,doerrGivingSoftwareIts2019,haferAssessingOpenSource2009,liSoftwareCitationReuse2016,liHowCitedResearch2017,parkResearchSoftwareCitation2019}, progress is being made, driven by different stakeholders:

\begin{itemize}
  \item Research software community initiatives such as the FORCE11 Software Citation Implementation Working Group (\href{https://www.force11.org/group/software-citation-implementation-working-group}{www.force11.org/group/software-citation-implementation-working-group}) build on established community standards~\cite{smithSoftwareCitationPrinciples2016} and bring together stakeholders to shape technical infrastructure and policy, and develop guidance~\cite{katzSoftwareCitationImplementation2019}; their activities concern the discursive and social functions directly, and the remaining functions indirectly.
  \item Domain, infrastructure and software communities develop software solutions for providing citation metadata~\cite{boettigerCitingPackages2012,yaroslavhalchenkoDuecreditDuecredit2019}, create repositories, information services, and indices~(e.g.,~\cite{githubMakingYourCode2016,shamirPracticesSourceCode2013,bonischSwMATHNewInformation2013,kingscollegelondonTERESAHToolsEregistry2014}), and develop metadata formats~\cite{druskat_stephan_2018_1405679,jonesCodeMetaExchangeSchema2017a}; their activities concern the context and social functions directly, and the remaining functions indirectly.
  \item Research policy researchers develop procedures for evaluating research software~\cite{gomez-diazEvaluationResearchSoftware2019}; their activities concern the social and compliance functions directly, and the remaining functions indirectly.
  \item Funding agencies update funding policies (cf.~\cite{piwowarAltmetricsValueAll2013}) and guidelines for scholarly practice~\cite{deutscheforschungsgemeinschaftdfgLeitlinienZurSicherung2019} to incorporate citation of research software; their activities concern the compliance function directly, and the remaining functions indirectly.
  \item Publishers establish editorial policies that require the citation of software, sometimes as a subset of data~\cite{parkResearchSoftwareCitation2019,hansonMakingDataMaximally2011}, or plan to do so~\cite{doerrGivingSoftwareIts2019}; their activities concern the context, social and reproducibility functions directly, and the remaining functions indirectly.
\end{itemize}

\noindent In this paper, I aim to contribute to the understanding of the requirements for the implementation of research software citation. To this end, I will investigate the output of the context function of citation, \emph{research citation graphs} (\emph{RCG}s), with the objective to answer the following research questions:

\begin{itemize}
  \item \textbf{RQ1:} What are the necessary changes in the model of research citation graphs to allow for the integration of research software, and the adoption of the citation functions?
  \item \textbf{RQ2:} What are the requirements for the implementation of software citation based on an updated model of research citation graphs?
  \item \textbf{RQ3:} What are current challenges for the instantiation of research citation graphs?
  \item \textbf{RQ4:} What applications do research citation graphs enable?
\end{itemize}

\section{Research citation graphs} 
\label{sec:research_citation_graphs}

Research products and the references between them can be modeled as a directed graph $G_1 = (V,E)$ where $V$ is a set of vertices
(or ``nodes''), and $E$ is a set of ordered pairs of nodes (i.e., ``directed edges''). The nodes in $V$ represent research products,
the edges represent reference relations (i.e., citation) between source nodes (the citing research product) and target nodes (the cited product).
This most basic model of a \emph{research citation graph} (\emph{RCG}) enables the context function of citation: It
helps understand what other research products a specific product relied on (``back-tracking''), or has led to (``forward-tracking''), in order to
build on this understanding in research, or conduct evaluations and measurements.
Both tracking methods can be implemented as graph traversal, where back-tracking follows outgoing edges,
and forward-tracking follows incoming edges.
The model also enables the social function of establishing trust and authority, where it is based on the citation of acknowledged trustworthy or authoritative research products rather than establishing references to their authors.

In order to fully exploit the social citation function, RCGs must model additional properties of research products:

\begin{itemize}
  \item The provision of academic credit requires the inclusion of author nodes, and authorship relations between them and research products, as does the establishment of trust and authority if it is based on individuals.
  \item Evaluation requires the inclusion of two classes of entity nodes: affiliations (research groups, institutions, countries, etc.), and respective affiliation relations between them and authors; ``product containers'' (journals, edited volumes, repositories, archives, etc.), and respective published-in \emph{part-of} relations between them and research products.
\end{itemize}

\noindent The model graph changes accordingly: Let $P$ be the set of all vertices $\{p_1,\ldots,p_n\}$ which represent research products, $A$ the set of all vertices $\{a_1,\ldots,a_n\}$ which represent authors, $I$ the set of all vertices $\{i_1,\ldots,i_n\}$ which represent evaluable author affiliations, and $C$ the set of all vertices $\{c_1,\ldots,c_n\}$ which represent entities which contain research products.
Let $\mathcal{V}$ be the set of disjoint sets $\{P,A,I,C\}$ of vertices in the RCG $G_2 = (V, E)$. Define $L : V \to \mathcal{V}$ to set

\begin{itemize}
  \item $L(v) = P$ when $v \in P \in \mathcal{V}$,
  \item $L(v) = A$ when $v \in A \in \mathcal{V}$,
  \item $L(v) = I$ when $v \in I \in \mathcal{V}$,
  \item $L(v) = C$ when $v \in C \in \mathcal{V}$.
\end{itemize}

\begin{figure}[!t]
  \centering
  \begin{tikzpicture}
      \node[state, very thick] (P1) [tape]{$\bm{p_{1}}$};
      \node[state] (A1) [below =of P1] {$a_{1}$};
      \node[state] (I1) [below =of A1] [rectangle]{$i_{1}$};
      \node[state] (A2) [right =of A1] {$a_{2}$};
      \node[state] (I2) [below =of A2] [rectangle]{$i_{2}$};
  
      \node[state, very thick] (P2) [right =of P1] [tape]{$\bm{p_{2}}$};
      \node[state] (A3) [right =of A2] {$a_{3}$};
      \node[state] (I3) [below =of A3] [rectangle]{$i_{3}$};
  
      \node[state, very thick] (P3) [right =of P2] [tape]{$\bm{p_{3}}$};
      \node[state] (A4) [right =of A3] {$a_{4}$};
      \node[state] (I4) [below =of A4] [rectangle]{$i_{4}$};
  
      \node[state] (C2) [above =of P3] [diamond]{$c_{2}$};    
      \node[state] (C1) [left= of C2] [diamond]{$c_{1}$};
  
      \path (P1) edge[very thick] node[el,above] {\scriptsize{\textbf{cite}}} (P2);
      \path (P1) edge[bend left=30, very thick] node[el,above] {\scriptsize{\textbf{cite}}} (P3);
  
      \path (A1) edge node[el,above] {\scriptsize{affil}} (I1);
      \path (A2) edge node[el,above] {\scriptsize{affil}} (I2);
      \path (A2) edge node[el,above] {\scriptsize{affil}} (I3);
      \path (A3) edge node[el,above] {\scriptsize{affil}} (I3);
      \path (A3) edge node[el,above] {\scriptsize{affil}} (I4);
      \path (A4) edge node[el,above] {\scriptsize{affil}} (I4);
  
      \path (P1) edge node[el,above] {\scriptsize{auth}} (A1); 
      \path (P1) edge node[el,above] {\scriptsize{auth}} (A2); 
      \path (P2) edge node[el,above] {\scriptsize{auth}} (A2); 
      \path (P2) edge node[el,above] {\scriptsize{auth}} (A3); 
      \path (P3) edge node[el,above] {\scriptsize{auth}} (A4); 
  
      \path (P1) edge node[el,above] {\scriptsize{pub-in}} (C1); 
      \path (P2) edge[bend right=40] node[el,above] {\scriptsize{pub-in}} (C1); 
      \path (P3) edge node[el,above] {\scriptsize{pub-in}} (C2); 
  \end{tikzpicture}
  \caption{Partial $G_2$-type RCG for a research product which references two other research products. The graph includes the respective $G_1$ graph (in bold print). \emph{Nodes}: $p_{n}$=research product,
  $a_{n}$=author, $i_{n}$=evaluable affiliation, $c_n$=evaluable publishing container; \emph{Edges}: \emph{cite}=``citation'' relation, \emph{auth}=``authored by'' relation, \emph{affil}=``affiliated with'' relation, \emph{pub-in}=``published in'' relation.}
  \label{fig:simple-graph}
  \end{figure}
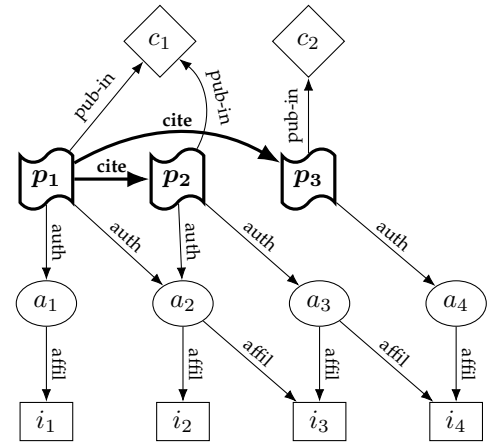

\noindent See Figure~\ref{fig:simple-graph} for an exemplary visualization.
In the past, the citation system, as the foundation of the academic credit and evaluation system, focused on journal articles, books, and conference papers~\cite{borgmanScholarshipDigitalAge2007}, and $G_2$-type RCGs are the output of the system's context function at this stage.
In addition to the more general reasons given above in section~\ref{sec:introduction}, research software must be integrated in the citation system also to realize the compliance and reproducibility functions of the current sociotechnical citation system.
Definitions of and guidelines for good scholarly practice change to reflect the digitalization of scholarship, and compliance with them requires research software to be cited. The recently updated guidelines for good scholarly practice by German funding agency Deutsche Forschungsgemeinschaft, for example, require that ``provenance of data, organisms, materials and software used in the research process is indicated and re-use documented; the original sources are cited''~\cite[p.~14, my translation]{deutscheforschungsgemeinschaftdfgLeitlinienZurSicherung2019}.
Computational reproducibility also naturally requires, amongst other information, e.g., about the computational environment, the correct identification of the software that has been used in published research~\cite{berez-kroekerReproducibleResearchLinguistics2018,alnoamanyComputationalReproducibilityResearcher2018}, and provision of this information through citation~\cite{doerrGivingSoftwareIts2019,smithSoftwareCitationPrinciples2016,stoddenEnhancingReproducibilityComputational2016,theyalelawschoolroundtableondataandcoresharingReproducibleResearch2010}.

In order to enable these functions in RCGs, they must model the citation-related specific properties which set software apart from other research products.
These properties relate to the software citation principles of \emph{specificity}, \emph{unique identification}, and \emph{attribution and credit}~\cite{smithSoftwareCitationPrinciples2016}.
Software differs from textual research products in the form its artifacts can take, in its notion of finality and the relationships between its artifacts, in the citability of its concepts, in its dynamicity, in the containment relationships between a product and its contributions, and in the roles which contribute to it.

Take an academic paper as example: As a research product, it is perceived to be a single artifact, available in a ``final'' version. This finality is a function of peer review, editorial acceptance, and ``adequate'' publication. Final published papers cannot have a new version after publication. Instead, any changes must again pass peer review, acceptance and publication, at which stage the changed paper is considered to be a discrete new research product.
Papers may also have non-final versions published as preprints.

In contrast, software development processes produce artifacts at different stages. Every commit (or ``revision'') in a version control system produces an artifact, a collection of source code and other files, and build processes may additionally produce one or more binary artifacts.
The notion of a ``final'' software product does not exist as such. This is due to the lack of a standardized publication process for software which is based on peer review. Instead, software can evolve over time, and any revision can be tagged as a version and ``released''. None of the versions can be considered ``final'', as at any time, any changes produce a newer version of the same software, not a discrete new research product.
Also, while versions of papers may or may not differ in content -- preprints may have the same content as a final publication -- versions of software will usually represent differing source codes.
When we talk about ``software'', we usually mean a version of a software that has been released or used.
Alternatively, ``software'' can also refer to the concept of a software (cf.~\cite{katzSoftwareCitationImplementation2019}), rather than a version, revision, or artifact: ``Microsoft Excel'' refers to the concept of a spreadsheet application, of which its versions are realizations.
Although a paper and its preprints are arguably also different realizations of the same concept, these paper concepts are never explicitly cited, whereas software concepts may be cited, e.g., in the case of pipelines or frameworks~\cite{katzSoftwareCitationImplementation2019}, or to understand the development
of a software in computer science research, or in software comparison.
In analogy, software concepts may be identifiable by unique identifiers as issued by repositories for digital research products such as Zenodo~\cite{nielsenZenodoNowSupports2017}.
For papers, identifiers for concepts are not issued across repositories. Usually, preprints and final publications will be archived on separate platforms.
Although preprint repositories such as arXiv do issue versioned identifiers, which point to the latest version when stripped of the version information, there are no cross-repository identifiers that can uniquely identify concepts of papers.

Defined as a ``set of instructions that direct a computer to
do a specific task''~\cite[p.~2]{chunSoftwarePersistenceVisual2005}, software is ``functionally active''~\cite[p.~2]{katzSoftwareVsData2016} (``dynamic''), i.e., it performs functions on data, whereas papers are clearly static. A software may have different states, and execute along
different paths at runtime. The final states and execution paths depend on configuration, interaction,
and possibly the data that is being processed. States
and execution paths define the actual ``dynamic product''
that is used to perform a specific software task.

A further difference between software and papers is the containment relationship between a product and the contributions to it.
Contributions to a research product can be \emph{active} or \emph{passive}, and \emph{direct}
or \emph{indirect}. In \emph{direct active contributions} to papers, contributors
influence the product directly through contributions of text, analyses,
ideas, etc. Direct active contributions to software can take the form of source
code, code comments, documentation, architectural design, API design, UI design, tests, code reviews, bug
reports, etc. With \emph{direct passive contributions}, a paper uses another product or parts thereof, by
building on it, refuting it, refining its analyses, contextualizing its findings, etc. Direct passive
contributions to software are mainly its ``dependencies'' -- i.e., other software -- but can also include other research products, such as
publications that describe algorithms, models or methods
implemented in the software.
\emph{Indirect contributions} to a
research product are direct or indirect contributions to
passive contributions to that product. Indirect software contributions to software are \emph{transitive dependencies}.

``Dependencies'' of a software $S$ are software components to which $S$ exhibits a degree of coupling.
If $S$ relies on functions of another software $S_1$, without which $S$ will not function as intended, $S_1$ is a dependency of $S$.
This usually means that $S$ calls functions from $S_1$, or uses its API in another way, e.g., through inheritance.
Dependencies can take different forms, as libraries, code fragments, or algorithms.
The defining quality of a dependency is that it is not part of the original, directly contributed, source code of a software.
Therefore, functions defined in a file $X$ that are called from functions defined in another file $Y$, are part of a dependency iff file $X$
is not part of the same codebase as file $Y$. This may include that file $X$ has other authors than file $Y$.
Original source code and dependency source code form the common codebase of a software.
At runtime, direct passive contributions (dependencies) and indirect passive contributions (transitive dependencies, i.e., the dependencies of direct dependencies) become part of the same ``software object'', as execution paths transcend boundaries between a software and its dependencies.
In contrast, direct passive contributions to papers are part of the product in the form of references;
indirect passive contributions are not part of a paper at all, but must instead be retrieved via backtracking traversal of an RCG.

Direct active contributions to papers are recognized through authorship, and direct passive contributions
through citation. Indirect contributions are not recognized, but can be discovered in RCGs.
Similarly, direct active contributions to a software should be recognized through authorship -- or acknowledged contributorship, see below -- whereas reliance on dependencies as direct passive contributions should
be recognized through citation.
This holds despite the \emph{part-of} relationship between dependencies, transitive dependencies, and the depending software at runtime,
when dependencies arguably become direct active contributions to a software product.
The software citation principles motivate this, by suggesting that software citation should generally address software source code~\cite{smithSoftwareCitationPrinciples2016}, which makes dependencies passive contributions.
Additionally, under a standard definition of authorship~\cite{internationalcommitteeofmedicaljournaleditorsICMJERecommendationsDefining}, the default recognition type for direct active contributions, authors of dependencies do not qualify for authorship of depending software: The contribution to a software through a dependency is substantial, but neither will they draft or revise the depending software, nor will they approve the version of the depending software to be published, or agree to be accountable for it. This categorically rules out authorship and motivates citation of dependencies instead.

Citation should attribute contributions to a research product to all contributors, and enable the provision of credit for a contribution.
There is increasing acknowledgment of the fact that direct contributions to research products can
take different forms than text production~\cite{brandAuthorshipAttributionContribution2015}, 
and that metadata should represent different
contribution types~\cite{mcnuttTransparencyAuthorsContributions2018}. 
This is the case for all types of research product, and specifically for software,
where creditable contributions greatly differ from those to papers, and include not just the writing of source code, but also
contributions to the architecture, design, documentation, engineering, management,
verification, validation, repair, maintenance, etc., of a software~\cite{chengActivityBasedAnalysisOpen2019,constantinouDevelopersExpertiseRoles2016,RecognizeAllContributors}. 
However, there is not yet a common understanding of which types of acknowledgable and creditable contributions there are across different types of research software.
It has also not yet been established whether there are qualitative differences between contribution types that may motivate a tiered concept of contributions, such as primary and secondary contributions. The schema.org ontology, for example, 
defines a term \emph{contributor} as ``a secondary contributor''~\cite{ContributorSchemaOrg}, whereas community initiatives such as All Contributors~\cite{RecognizeAllContributors} suggest to reward ``every contribution, not just code'' without defining differently valued contributions.
It is therefore not possible to add contributor types to a model of RCGs at this stage in a meaningful and future-proof way.
Instead, different stakeholders -- software producers, researchers, publishers, institutions, policy makers, and others --  should collaborate to develop and implement policies and ontologies that allow for a more differentiated model of author- and contributorship across all acknowledgable research products.
This model should replace the traditional authorship model and be reflected in the metadata for research products that is provided at publication, for purposes of citation, credit, etc.

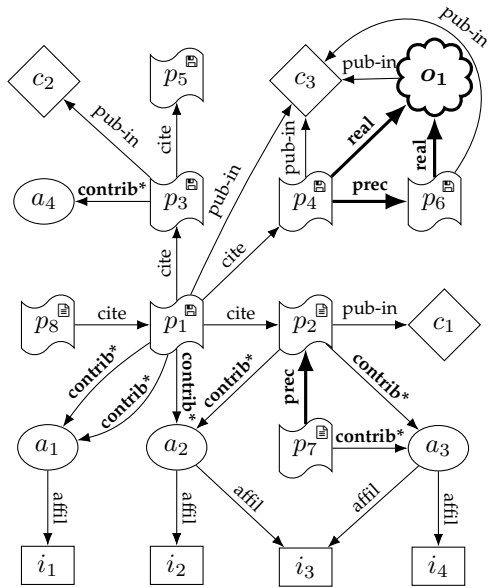
\begin{figure}[!t]
  \centering
  \begin{tikzpicture}
      \node[state] (P8) [tape]{$p_{8}$};
      \node[state] (P1) [right =of P8] [tape]{$p_{1}$};
      \node [left =of P1, xshift=0.05cm, yshift=0.2cm] {\tiny{\faFileTextO}}; 
      \node [right =of P8, xshift=0.35cm, yshift=0.2cm] {\tiny{\faFloppyO}}; 
      \node[state] (P2) [right =of P1] [tape]{$p_{2}$};
      \node [right =of P1, xshift=0.38cm, yshift=0.2cm] {\tiny{\faFileTextO}}; 
      \node[state] (P3) [above =of P1] [tape]{$p_{3}$};
      \node [above =of P1, xshift=0.2cm, yshift=0.35cm] {\tiny{\faFloppyO}}; 
      \node[state] (P4) [right =of P3] [tape]{$p_{4}$};
      \node [right =of P3, xshift=0.35cm, yshift=0.2cm] {\tiny{\faFloppyO}}; 
      \node[state] (P5) [above =of P3] [tape]{$p_{5}$};
      \node [above =of P3, xshift=0.2cm, yshift=0.35cm] {\tiny{\faFloppyO}}; 
      \node[state] (P6) [right =of P4] [tape]{$p_{6}$};
      \node [right =of P4, xshift=0.35cm, yshift=0.2cm] {\tiny{\faFloppyO}}; 
      \node[state] (P7) [below =of P2] [tape]{$p_{7}$};
      \node [below =of P2, xshift=0.22cm, yshift=0.1cm] {\tiny{\faFileTextO}}; 
  
      \node[state] (A4) [left =of P3] {$a_{4}$};
      \node[state] (A1) [below =of P8] {$a_{1}$};
      \node[state] (A2) [below =of P1] {$a_{2}$};
      \node[state] (A3) [right =of P7] {$a_{3}$};
  
      \node[state, very thick] (O1) [above =of P6, yshift=-0.2cm] [cloud, align=center]{$\bm{o_{1}}$};
  
      \node[state] (C1) [right= of P2] [diamond]{$c_{1}$};
      \node[state] (C2) [above= of A4, yshift=-0.2cm] [diamond]{$c_{2}$};
      \node[state] (C3) [above= of P4] [diamond, yshift=-0.2cm]{$c_{3}$};
  
      \node[state] (I1) [below =of A1] [rectangle]{$i_{1}$};
      \node[state] (I2) [below =of A2] [rectangle]{$i_{2}$};
      \node[state] (I3) [below =of P7] [rectangle]{$i_{3}$};
      \node[state] (I4) [below =of A3] [rectangle]{$i_{4}$};
  
  
      \path (P8) edge node[el,above] {\scriptsize{cite}} (P1);
      \path (P1) edge node[el,above] {\scriptsize{cite}} (P2);
      \path (P1) edge node[el,above] {\scriptsize{cite}} (P3);
      \path (P1) edge node[el,above] {\scriptsize{cite}} (P4);
      \path (P3) edge node[el,above] {\scriptsize{cite}} (P5);
  
      \path (P1) edge[bend left=30] node[el,above] {\scriptsize{\textbf{contrib*}}} (A1); 
      \path (P1) edge[bend right=10] node[el,above] {\scriptsize{\textbf{contrib*}}} (A1); 
      \path (P1) edge node[el,above] {\scriptsize{\textbf{contrib*}}} (A2); 
      \path (P2) edge node[el,above] {\scriptsize{\textbf{contrib*}}} (A2); 
      \path (P2) edge node[el,above] {\scriptsize{\textbf{contrib*}}} (A3); 
      \path (P3) edge node[el,above] {\scriptsize{\textbf{contrib*}}} (A4); 
      \path (P7) edge node[el,above] {\scriptsize{\textbf{contrib*}}} (A3); 
  
      \path (P7) edge[very thick] node[el,above] {\scriptsize{\textbf{prec}}} (P2); 
      \path (P4) edge[very thick] node[el,above] {\scriptsize{\textbf{prec}}} (P6); 
  
      \path (P4) edge[very thick] node[el,above] {\scriptsize{\textbf{real}}} (O1);     
      \path (P6) edge[very thick] node[el,above,xshift=-0.1cm,yshift=0.02cm] {\scriptsize{\textbf{real}}} (O1);    
  
      \path (P2) edge node[el,above] {\scriptsize{pub-in}} (C1); 
      \path (P3) edge node[el,above] {\scriptsize{pub-in}} (C2); 
      \path (P4) edge node[el,above] {\scriptsize{pub-in}} (C3); 
      \path (P6) edge[out=55, in=45, looseness=2.05] node[el,above] {\scriptsize{pub-in}} (C3); 
      \path (O1) edge node[el,above] {\scriptsize{pub-in}} (C3); 
      \path (P1) edge[bend left=3] node[el,above] {\scriptsize{pub-in}} (C3); 
  
      \path (A1) edge node[el,above] {\scriptsize{affil}} (I1);
      \path (A2) edge node[el,above] {\scriptsize{affil}} (I2);
      \path (A2) edge node[el,above] {\scriptsize{affil}} (I3);
      \path (A3) edge node[el,above] {\scriptsize{affil}} (I3);
      \path (A3) edge node[el,above] {\scriptsize{affil}} (I4);

  
  

  \end{tikzpicture}
  \caption{Partial $G_3$-type RCG for an exemplary textual research product ($p_8$) which cites an exemplary software ($p_1$) with two software dependencies ($p_3, p_4$), and a citation to a paper ($p_2$). Additional and changed elements in comparison to the $G_2$ model in bold print. Software products marked with \faFloppyO, textual products marked with \faFileTextO. \emph{Nodes}: $p_{n}$=\textbf{version} of a research product,
  $a_{n}$=\textbf{acknowledged contributor}, $i_{n}$=evaluable affiliation, $c_n$=evaluable publishing container; \emph{Edges}: \emph{cite}=``citation'' relation, \emph{contrib*}=``contributed to by'' relation, \emph{affil}=``affiliated with'' relation, \emph{pub-in}=``published in'' relation, \emph{prec}=``precedes'' relation, \emph{real}=``realizes'' relation.}
  \label{fig:complex-graph}
  \end{figure}

In summary, the described specific properties of software yield new requirements for a $G_3$ RCG model which integrates software research products, and supports the compliance and reproducibility functions:


\begin{itemize} 
  \item The versionability of software (and other) research products requires no new model elements, but the re-definition of the set $P$ as the set of all vertices $\{p_1,\ldots,p_n\}$ which represent \emph{versions} of research products.
  \item The specific relationship between versions as realizations of concepts and the respective concepts requires the addition of concept nodes, i.e., a new set $O \in \mathcal{V}$, which is the set of all vertices $\{o_1,\ldots,o_n\}$ which represent \emph{concepts} of research products. Alternatively, $P$ could be re-defined as the set of all vertices $\{p_1,\ldots,p_n\}$ which represent \emph{versions or concepts} of research products. For reasons of clarity, I will pursue the first option in the following.
  \item The relationships between versions and concepts also require the addition of realization relations, i.e., edges from nodes in $P$ to nodes in $O$.
  \item For cases where concepts that overarch versions of a research product remain unidentifiable -- as is usually the case with papers -- the relationships between versions of a research product require the addition of order relations, i.e., edges from nodes in $P$ to other nodes in $P$ which define one version (the source node) as the predecessor of another version (the target node). These edges allow for the analysis of cumulative impact over versions of a research product.
  \item A differentiated model of contributorship to research products, and specifically software, requires the addition of edges representing different contribution types. As this is currently not possible in lieu of an agreed-upon model of acknowledgable contributorship, the relations formerly denoting authorship have been labeled \emph{contrib*} in Figure~\ref{fig:complex-graph}, instead of \emph{auth}, where the label is to be understood to reflect different types of contributions, including both traditional authorship and the more fine-grained contributions to digital research products.
\end{itemize}

\noindent These changes take into account almost all of the specific properties of software as a research product.
``Dynamic products'' have not been included here because they represent objects different from the principle target of software citation, i.e., source code (cf.~\cite{smithSoftwareCitationPrinciples2016}). They will be discussed further in section~\ref{sec:challenges_for_the_instantiation_of_rcgs}.
See Figure~\ref{fig:complex-graph} for a visualization showing an example of the updated $G_3$ model for RCGs. The figure shows a paper $p_8$, which cites a software version $p_1$. $p_1$ has been contributed to by two contributors, one of which ($a_1$) has made contributions of two different types, e.g., writing source code and writing documentation. $p_1$ also cites two dependencies $p_3$ and $p_4$. $p_3$ cites its own dependency $p_5$, making $p_5$ a transitive dependency of $p_1$. $p_4$ is a realization of software concept $o_1$, and precedes another version of the same software, $p_6$, which is also a realization of $o_1$. $p_1$ also cites a paper $p_2$, e.g., describing an algorithm which $p_1$ implements. $p_2$ in turn has a predecessor in the preprint $p_7$, with which it shares one contributor. For $p_2$ and $p_7$, no concept has been published.

Taken together, the updated $G_3$ model for research citation graphs enables all functions of the current sociotechnical system of citation for research software (RQ1).
With regards to the discursive function, some systemic limitations apply. First of all, the acknowledgement of direct passive software contributions to a research software, or failure to do so, can only
be used to shape and organize discourses of scholarly credibility, authority, and relevance to a certain extent. As dependencies are hard-wired
into the software product, only the preference of dependencies over functionally equivalent others before or while creating the product have the potential to affect a given discourse.
To neglect, or favour, a specific dependency over another usually means a trade-off of functionality for a research software. These choices will be
determined by the functional and/or engineering needs of any given software project, and are not likely to be used as discursive actions.
A ``re-writing of the past'', by wilfully neglecting to \emph{reference} specific dependencies and their import in the scholarly discourse, becomes impossible
as soon as they are \emph{used}, and thus become an inseparable part of, a software.
Some build systems have a concept of optional dependencies. Optional dependencies, i.e., those that can optionally be used in a software which also defines a fallback mechanism for the case that the optional dependency cannot be resolved and used on a given system, may be more prone to allow discursive actions, but the discussion of this corner case is out of scope for this paper.
Another case where the discursive function may be of interest is in references from a software to a text-based research product. This could be the case
when a software implements an algorithm described in a paper, and the paper should be cited to enable due credit for its contributors. Choosing to not cite
the paper when it should be, is arguably discursive action with respect to creditability and relevance of research products. On the other hand, this goes
against currently promoted good scholarly practice~\cite{deutscheforschungsgemeinschaftdfgLeitlinienZurSicherung2019}, and presents another corner case.
Finally, and most importantly, the discursive function does not influence RCGs directly in terms of elements in a graph. Rather, its output must be studies
on what is not in a given RCG instance when it should be, or what is in a graph when it should not be.

Based on the discussion and development of a useful model of research citation graphs that integrate software in this section, I will discuss current challenges for the instantiation of RCGs in the following section.


\section{Challenges for the instantiation of RCGs} 
\label{sec:challenges_for_the_instantiation_of_rcgs}


Research citation graphs have a number of potential applications, which I will discuss in section~\ref{sec:applications_for_research_citation_graphs}.
A prerequisite for these applications is the instantiation of research citation graphs for actual research products, including software.
RCGs are built from distributed metadata, probably represented as linked data.
An example of this is CodeMeta~\cite{jonesCodeMetaExchangeSchema2017a}, a linked data exchange schema for software metadata which extends the schema.org vocabulary and is implemented in JSON-LD. CodeMeta files can, for example, represent software as well as other research products as references of a software. For journal papers, JATS~\cite{beckNISOZ39962011} can be used to retrieve metadata and resolve references.
Instantiated RCGs can, e.g., be stored in (graph) databases, represented in main memory, or visualized as graphs.

In theory, RCGs are instantiated by recursive resolution of references from a ``root'' research product to retrieve the set $P \in \mathcal{V}$ for a graph $G = (V, E)$; for each $p \in P$, retrieval of the product-specific metadata to retrieve the sets $A$, $I$, $C$, and $O$ for $G$; deduplication of the vertices in $V$ for $G$.

This process yields requirements for an implementation of software citation (RQ2), which also reflect the software citation principles~\cite{smithSoftwareCitationPrinciples2016} (in parentheses):

\begin{itemize}
  \item Publications and publication metadata -- including references -- are available digitally, and metadata are machine-actionable (``Persistence'', ``Accessibility'');
  \item research products duly and correctly cite references including software in publications (``Importance'', ``Specificity'', ``Credit and attribution'');
  \item publications are uniquely identifiable through a machine-actionable identifier (``Accessibility'', ``Unique identification'');
  \item contributors, research institutions and other evaluable entities as well as publication platforms can be deduplicated, i.e., are uniquely identifiable.
\end{itemize}

\noindent To meet these requirements for research software, a number of challenges (RQ3) have to be overcome first.
The publication process for textual research products is well-established and involves peer review, editorial acceptance and adequate publication together with curated and complete metadata and unique identification to enable the citation use case.
No such process is yet in place for software, although software journals such as JORS~\cite{JournalOpenResearch} or JOSS~\cite{smithJournalOpenSource2018} aim to provide a similar workflow, but do not meet the ``Importance'' principle~\cite{smithSoftwareCitationPrinciples2016}, as they publish metapapers, and not the software itself. Additionally, while references to text publications are recorded in their metapapers, software references (dependencies) are not. Alternatives include automated deposits from source code repositories such as GitHub to general purpose archives such as Zenodo, which provide unique identification, but do not require or curate citation-relevant metadata, including references. Similarly, Software Heritage~\cite{abramaticBuildingUniversalArchive2018} harvests source code and provides unique identification, but does not require citation-relevant metadata, including references.

In order to establish a publication process for software similar to that of papers, action is required from many different stakeholders. Research software creators will have to embed publication steps into their workflows, which also reflect short, iterative cycles of software development. These publication steps should be supported by repositories and archives, as well as publishers, which need to adopt and process metadata schemas that include references to other research products, and make reference metadata available in addition to product metadata~\cite{shottonSettingOurBibliographic2015}.
Software publications must further be considered in evaluation processes by research institutions and funders, and thereby creating incentives for creators and institutions to publish research software.
It will also have to be determined at which stage peer review and editorial acceptance -- including the required curation of metadata -- may be integrated. Peer review, editorial oversight, and metadata curation for software may be in the scope of publishers collaborating with archives. Instead of publishing metapapers, software journals could manage the peer review and acceptance process for a software deposit in archives. Suitable business models for this will have to be established, especially as software products can usually be published in cycles much shorter than those for papers. Another option for the curation of metadata would be for libraries to engage and adopt this task, again in collaboration with archives and research data management. Suitable metadata and exchange formats exist\cite{druskat_stephan_2018_1405679,jonesCodeMetaExchangeSchema2017a}.

Such adequate publication practices for software may also support the due citation of research software, which is still not widely established~\cite{haferAssessingOpenSource2009,doerrGivingSoftwareIts2019,howisonSoftwareScientificLiterature2016,liSoftwareCitationReuse2016,liHowCitedResearch2017,parkResearchSoftwareCitation2019}. In order to establish due citation of software, a culture change needs to take place across research, which can be driven from two directions: top-down by policy makers, institutions, funders, publishers and publishing platforms which can require the citation of software in research, and the provision of suitable metadata for software publications, and reward it based on adapted evaluation practices; bottom-up by educators, peer reviewers, editors, researchers, research software engineers, etc., which educate about, insist on, enable, and practice due citation of software and provision of citation metadata for software.

The unique identification of software products can be achieved through publication via archives that provide DOIs or similar persistent identifiers. While this solves the technical side of unique identification of software, the cultural challenge, i.e., the adoption of respective publication practices, remains to be solved as discussed above.

The unique identification of contributors and institutions can be achieved through the use of identifiers such as ORCID. Again, this solves the technical side, but adoption remains a cultural challenge. This can be tackled through encouragement, request and requirement of adoption of identifiers from funders, publishers, and institutions, as well as through education and exemplary practice by researchers, software creators, and institutions.

While the requirements discussed above apply to all types of research products, there are two particular aspects of research software which further affect the instantiation of RCGs.
As mentioned in section~\ref{sec:research_citation_graphs}, software has two states: the static state of its source code, and the dynamic state at runtime.
While strictly speaking, the dynamic state is irrelevant in a discussion of software citation as defined by the software citation principles~\cite{smithSoftwareCitationPrinciples2016}, there are solutions that may enable the instantiation of RCGs constricted to a single software at runtime.
Software that documents the executed paths taken in a complex software product to produce a research outcome at runtime can potentially produce an RCG for this execution. Duecredit~\cite{yaroslavhalchenkoDuecreditDuecredit2019} implements this concept for software written in Python. It registers references for portions of code at the module and function levels, and can inject references for dependencies. Its output is a list of references that represent the code that has been called during execution. This output can potentially be transformed to an RCG, albeit a local one which will usually contain only first and second level references of a software, depending on the downstream provision of the respective metadata by dependency projects. Additionally, the manually provided reference metadata cannot be verified.

While duecredit may not be suitable for the instantiation of non-local, larger-scale static RCGs, it brings into focus the fact that research software will
often include dependencies or transitive dependencies that are not research products, and which will therefore not be published in an adequate way, and not come with citation-relevant metadata.
In order to provide relevant context for research products, and enable all functions of citation for research software, these ``hidden'' contributions to research must be included in RCGs. This touches, in fact, the core of the discursive function of citation~\cite{foucaultArchaeologyKnowledge1982}.
In lieu of publications, unique identifiers, and curated metadata for non-research software, this can be done by applying software engineering methods.
Through static code analysis using manifests, build configurations, or import statements in conjunction with repository mining methods, a dependency graph can be retrieved for a given software, which can be transformed into a partial RCG for research software dependencies for which no machine-actionable metadata or unique identifiers are provided. I will investigate this method in future research.

In summary, the feasibility of instantiating RCGs that include research software is currently limited. This is due to unsatisfactory software publication practices, lack of provided correct and complete metadata, and insufficient software citation practices induced by lack of incentive and requirement to cite software. Solutions are being proposed and developed in technology~\cite{druskat_stephan_2018_1405679,jonesCodeMetaExchangeSchema2017a}, policy~\cite{deutscheforschungsgemeinschaftdfgLeitlinienZurSicherung2019}, theory~\cite{smithSoftwareCitationPrinciples2016} and other areas~\cite{katzSoftwareCitationImplementation2019}. These solutions support a culture change towards software citation implementation.
Progress in software citation implementation will gradually unlock applications for RCGs. I will provide an outlook on exemplary applications in the following section.


\section{Applications for research citation graphs} 
\label{sec:applications_for_research_citation_graphs}

RCGs enable different analyses of the context of research products (RQ4).
In this section, I will outline potentially useful analyses based on the visualization of the $G_3$ model for RCGs in Figure~\ref{fig:complex-graph}.
The exemplary analyses also serve as indirect evaluation of the model.

\textbf{Back-tracking exploration} The context of research products can be explored using RCGs to find out which preceding research a research product builds on.
This can be done by traversing the graph starting from $p_8$ and following outgoing edges of type \emph{cite}.
The analysis shows for example, that the paper $p_8$ indirectly builds on research published in $p_2$.
The implementation of software citation solicited above makes this insight possible, as without citation of $p_1$ in $p_8$, and citation of $p_2$ in $p_1$, the
relation between $p_8$ and $p_2$ would remain hidden.

\textbf{Citation tracking} The context of research products can be explored using RCGs to find out which research builds upon a given research product.
This can be done, e.g., by traversing the graph starting from a given research product node and following incoming edges of type \emph{cite}.
The analysis shows, for example, that paper $p_2$ has been cited by software $p_1$, in addition to any other papers citing it (not visualized).
Again, software citation makes this insight possible by providing not only references \emph{to} software, but also references of other products \emph{from} software.

\textbf{Tracking of concept citation} $G_3$-type RCGs enable citation analyses for concepts in addition to products.
This can be done by traversing the graph starting at $o_1$, following incoming nodes of type \emph{real} to products realizing the concept published in $o_1$, and
consecutively following outgoing \emph{cite} relations from the realizing products.
Given the citation of $p_4$ in $p_1$, and assuming that $p_6$ was cited in another product $p_9$ (not visualized), this analysis yielded a citation count of 2 for
the concept of a software which has been realized in two implementation versions of the software.

\textbf{Contribution role analysis} Traversing the graph starting from $p_1$ and following outgoing \emph{contrib*} relations allows an analysis of how
roles are distributed over contributors to a research product.
Once a sufficient model for contributions to research has been established, this also allows for a fine-grained evaluation of contributors with respect
to their skill sets and creditable contributions.

\textbf{Self-citation analyses} RCGs enable self-citation analyses by finding nodes in $A$ (for contributor-based analyses) or $C$ (for, e.g., journal-based analyses) with more than one incoming relation of type
\emph{contrib*}, and looking at whether their source nodes are connected directly with a \emph{cite} relation.

\textbf{Analysis of software development practices} Traversing the graph from a software product node, e.g., $p_6$, following incoming (or outgoing) \emph{prec} relations, creates a timeline of versions of a software. Provided the respective publication dates deposited in the respective machine-readable metadata (not visualized), the common target node of the \emph{real} relations can be taken into account to analyse the software development model employed for the implementation of software concepts, e.g., $o_1$. Given short timespans between versions, for example, an agile process could be assumed.

\textbf{Credit for ``hidden'' contributions to research} Assuming that $p_5$ is a commercially-developed software which was never intended to be published as a research product, and given that $p_1$ is research software, traversing the graph from $p_1$ following outgoing \emph{cite} relations enables attribution of
and credit for the contributors to $p_5$ (not visualized) for their contribution to the research published as $p_1$.
One obvious challenge here is the retrieval of complete and correct contributor information so that contributors to $p_5$ are correctly attributed and can receive credit.

\textbf{Retrieval of transitive credit} The last exemplary analysis (\emph{Credit for ``hidden'' contributions to research}) already hints at the usability of RCGs for calculating fractional credit to be recorded in transitive credit maps for a research product. When fractional credit for a research product is not only distributed over authors -- as is currently done implicitly through the order of authors lists for papers -- but also over contributors and cited research products, a complete credit map for a research product is created. A credit map for a product A also feeds into the credit map of a product B that cites A. The main principle of transitive credit~\cite{katzTransitiveCreditMeans2014,katzTransitiveCreditJSONLD2015} is that a contributor to product A can therefore receive credit for product B. If a fractional credit weight is determined for all contributors and references (i.e., built-upon research products), the credit weight for a single contributor or cited product can be determined transitively. If a software S is jointly developed by two people who share credit equally, both receive $.5$ fractional credit of the summary credit of $1$ that can be distributed for any given research product. If S is cited in a paper P, and the fractional credit for S for its contribution to P is weighted to $.2$ (of $1$), each contributor to S receives fractional credit of $.1$ transitively for P.

Different systems of weighting have been suggested for fractional credit, such as contribution taxonomies~\cite{brandAuthorshipAttributionContribution2015,RecognizeAllContributors} which could define default credit weights for different contribution types as templates for fractional credit. The same could be done for dependencies, which could be assigned a fixed (small) weight template for contributions to another software.
However, this would disregard the fact that, for example, a software A has a different weight for a software B which provides a wrapper API for A, than for a software C which uses a single function of A and has a large number of additional dependencies.
In future research, I will develop a weighting system of
fractional credit for software dependencies which instead is based on software engineering metrics such as function call frequencies and complexity. In contrast to the retrieval of credit weights for other products than software, such a
system does not rely on access to publication metadata, as it can use software engineering artifacts such as manifests and build configurations.
It thus exploits the actual conditions found in the software and dependencies, rather than a meta-representation.
The
fractional credit weights will still need to be registered for publications, which requires solutions to challenges described in section~\ref{sec:challenges_for_the_instantiation_of_rcgs}.

Once fractional credit values are registered in metadata for research products during publication, RCGs make it easy to ascribe transitive credit to contributors to cited references, including dependencies. RCGs embed dependencies on a par with other research products, following the software citation principles of ``Importance''~\cite{smithSoftwareCitationPrinciples2016}, while preserving the actual reference chains between software and its dependencies. In contrast, alternatives to the representation of
dependencies for credit in RCGs (cf.~\cite{ahaltNSFWorkshopSupporting}) may obfuscate these chains of references by citing software more granularly from a research product, i.e., software and dependencies are cited at one and the same depth rather than at the depth they have in the actual dependency graph; or, they place acknowledgable contributions outside of citations altogether, and instead provide provenance information on software websites for example, thereby disregarding the ``Importance'' principle~\cite{parkResearchSoftwareCitation2019}.


\section{Conclusion} 
\label{sec:conclusion}
In this paper, I have introduced a directed graph model for research citation graphs that integrates software and dependencies (RQ1).
This model can be used to determine requirements for the implementation of software citation based on established principles~\cite{smithSoftwareCitationPrinciples2016} (RQ2).
These requirements are not currently met, and the current state of software citation poses challenges for the instantiation of the model (RQ3). These include:
a lack of standardized publication practices for software; insufficient metadata provision and curation practices;
a lack of incentives to cite software and give due credit to contributors to research software; insufficient use of
unique identifiers for researchers and institutions.
Some of these challenges can be tackled with software engineering methods and the application of good scholarly practice, others rely on a culture change concerning the attitude to software as a research product, and the implementation of respective practices.
Once these challenges are overcome, research citation graphs based on the presented model enable a number of useful
applications (RQ4), such as bibliometric and scientometric studies, analyses of software development workflows applied in research,
and transitive credit.

\ifCLASSOPTIONcompsoc
  \section*{Acknowledgments}
\else
  \section*{Acknowledgment}
\fi
I would like to thank the discussion group on citation and
rewarding systems at the Workshop on Sustainable Software
Sustainability 2019 on 25 April 2019 in The Hague, Netherlands
(\href{https://www.software.ac.uk/wosss19}{www.software.ac.uk/wosss19}).
Discussion within the group has helped me to better understand the context for embedding software in the citation graph of
research. The members of this group were: Neil Chue Hong, Gerard Coen, James
Davenport, Leyla Garcia, Robert Haines, Catherine Jones, Adriaan Klinkenberg,
Rachael Kotarski, Mateusz Kuzak, Brett Olivier, Esther Plomp, Shoaib Sufi,
Stephanie van de Sandt, and Bettine van Willigen.
I would also like to thank three anonymous reviewers for their very helpful suggestions.
\ifCLASSOPTIONcaptionsoff
  \newpage
\fi

\bibliographystyle{IEEEtran}
\bibliography{IEEEabrv,2019-cise-software-citation}

\begin{IEEEbiography}[{\includegraphics[width=1in,height=1.25in,clip,keepaspectratio]{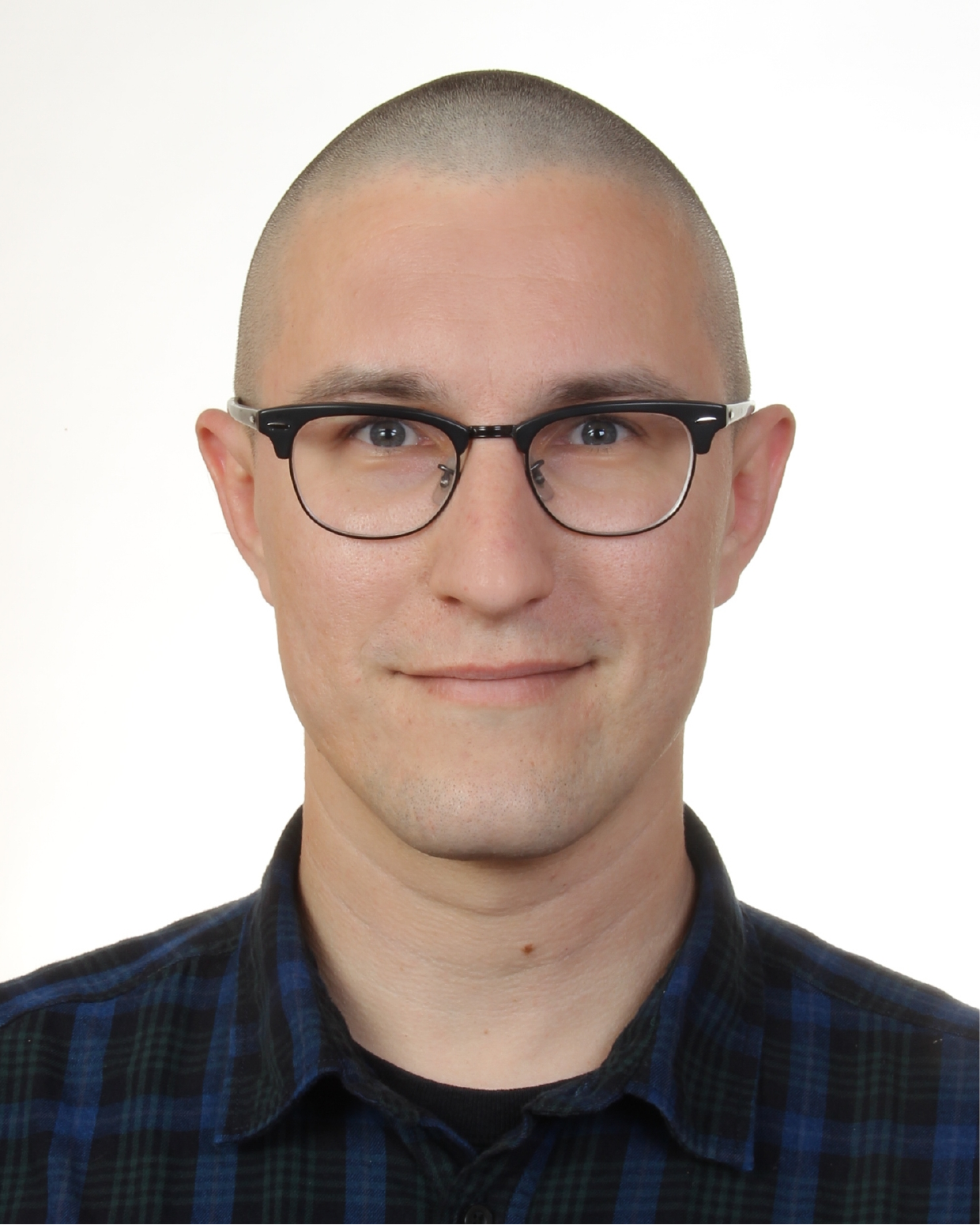}}]{Stephan Druskat} holds an MA in English, Modern German Literature and Linguistics from the Free University of Berlin, Germany.
  He is a Research Software Engineer, working in linguistics, and a PhD candidate in Software Engineering at the German Aerospace Center (DLR) and the Computer Science Department at Humboldt-Universität zu Berlin in Berlin, Germany. In his work, he focuses on research software sustainability and software citation. He is a Special Collaborator of the Software Sustainability Institute (UK), and a board member of de-RSE e.V. - Society for Research Software (Germany).
\end{IEEEbiography}
\end{document}